\documentclass[conference]{IEEEtran}
\IEEEoverridecommandlockouts
\usepackage{cite}
\usepackage{amsmath,amssymb,amsfonts}
\usepackage{algorithmic}
\usepackage{graphicx}
\usepackage{textcomp}
\usepackage{xcolor}
\def\BibTeX{{\rm B\kern-.05em{\sc i\kern-.025em b}\kern-.08em
    T\kern-.1667em\lower.7ex\hbox{E}\kern-.125emX}}

\usepackage{multirow}

\usepackage{hyperref}
\usepackage{algorithm}
\usepackage{algorithmic}
\usepackage{comment}
\usepackage{todonotes}
\usepackage{svg}

\begin{document}

\title{Towards Effective Authorship Attribution: Integrating Class-Incremental Learning}

\author{
\IEEEauthorblockN{Mostafa Rahgouy}
\IEEEauthorblockA{Auburn University\\
MZR0108@auburn.edu}
\and

\IEEEauthorblockN{Hamed Babaei Giglou}
\IEEEauthorblockA{TIB Leibniz Information Centre for\\ Science Technology \\
hamed.babaei@tib.eu}
\and

\IEEEauthorblockN{Mehnaz Tabassum}
\IEEEauthorblockA{
Auburn University\\
mzt0078@auburn.edu}
\and

\IEEEauthorblockN{Dongji Feng}
\IEEEauthorblockA{Gustavus Adolphus College\\
djfeng@gustavus.edu}
\and

\IEEEauthorblockN{Amit Das}
\IEEEauthorblockA{Auburn University\\
azd0123@auburn.edu}
\and

\IEEEauthorblockN{Taher Rahgooy}
\IEEEauthorblockA{Meta\\
trahgooy@meta.com}
\and

\IEEEauthorblockN{Gerry Dozier}
\IEEEauthorblockA{Auburn University\\
doziegv@auburn.edu}
\and

\IEEEauthorblockN{Cheryl D. Seals}
\IEEEauthorblockA{Auburn University\\
sealscd@auburn.edu}
}

\maketitle

\begin{abstract}
Authorship Attribution (AA) is the process of attributing an unidentified document to its true author from a predefined group of known candidates, each possessing multiple samples. The nature of AA necessitates accommodating emerging new authors, as each individual must be considered unique. This uniqueness can be attributed to various factors, including their stylistic preferences, areas of expertise, gender, cultural background, and other personal characteristics that influence their writing. These diverse attributes contribute to the distinctiveness of each author, making it essential for AA systems to recognize and account for these variations. However, current AA benchmarks commonly overlook this uniqueness and frame the problem as a closed-world classification, assuming a fixed number of authors throughout the system's lifespan and neglecting the inclusion of emerging new authors. This oversight renders the majority of existing approaches ineffective for real-world applications of AA, where continuous learning is essential. These inefficiencies manifest as current models either resist learning new authors or experience catastrophic forgetting, where the introduction of new data causes the models to lose previously acquired knowledge. To address these inefficiencies, we propose redefining AA as Class-Incremental Learning (CIL), where new authors are introduced incrementally after the initial training phase, allowing the system to adapt and learn continuously. To achieve this, we briefly examine subsequent CIL approaches introduced in other domains. Moreover, we have adopted several well-known CIL methods, along with an examination of their strengths and weaknesses in the context of AA. Additionally, we outline potential future directions for advancing CIL AA systems. As a result, our paper can serve as a starting point for evolving AA systems from closed-world models to continual learning through CIL paradigms.

\end{abstract}


\section{Introduction}
Authorship attribution (AA)~\cite{AA06} has a rich history in natural language processing (NLP) and has been extensively reviewed by researchers~\cite{koppel2009computational, stamatatos2009survey, Coyotl06}, covering cases with a minimal number of authors to scenarios involving hundreds of thousands of authors~\cite{blog100, Abbasi2022}. The analysis ranges from short texts with limited content, such as social media tweets~\cite{twitteraa}, to lengthy documents like books~\cite{tripto2023word2vec}, spanning from a casual tone to a more formal one~\cite{corbara2023same}. Such efforts have been undertaken to address the diversity of the real-world applications of AA, including (I) \textit{Digital forensics}~\cite{forensic14}, which addresses crimes related to technology misuse (e.g., fraud, harassment, fake news dissemination, or the creation of fictitious suicide notes). (II) \textit{Academic settings}, by aiding in plagiarism detection and contributing to copyright enforcement. (III) \textit{Software engineering}, where it is utilized for tasks such as source code AA~\cite{codesurvey20} and the identification of malicious software~\cite{MalwareAA23}. However, despite the progress made in bridging research with real-world AA applications, a fundamental principle has been overlooked. Prior works have not adequately considered the open-world nature of the AA task, wherein systems must accept and continuously learn from new and emerging authors. This oversight arises because AA is often formulated as a closed-world classification task, akin to other NLP tasks such as sentiment analysis~\cite{medhat2014sentiment} or text classification~\cite{kowsari2019text}, where the number of classes typically remains static. This can present a serious challenge given that most AA applications operate in a streaming fashion. For example, in a plagiarism detection system for academic assignments~\cite{homework16}, new authors (students) are continually added over time (e.g., with each new semester the course is offered), necessitating systems that can adapt and learn in this ongoing, dynamic manner. Similar demands are observed in other AA tasks, such as scholarly plagiarism detection and malicious software identification. This adaptability is crucial for maintaining the effectiveness and accuracy of AA applications in dynamic and evolving environments. 

To address the aforementioned challenge, we propose the Class-Incremental Learning (CIL)~\cite{TIAN2024307} paradigm in AA. In this paradigm, the model is encouraged to retain the knowledge it has already acquired while also being capable of learning about new authors as they are introduced. Consequently, our contributions are as follows:
\begin{itemize}
    \item We introduced the CIL paradigm in AA, which, to the best of our knowledge, is the first of its own kind.
    \item We made our implementation publicly available at GitHub repository~\footnote{\href{https://github.com/MostafaRahgouy/AA-CIL}{https://github.com/MostafaRahgouy/AA-CIL}} for the research community.
    \item We present a brief and high-level taxonomy of existing CIL approaches in other domains, discussing their strengths and weaknesses regarding adaptation to the AA. 
    \item We implement several well-known CIL models and evaluate them on widely used AA datasets.
    \item Finally, we outline potential future directions for the research community to explore this emerging task.
\end{itemize}

\section{Background}
In this section, we will provide background on CIL in the context of AA by presenting our formulation of AA using CIL. Next, we will explore the landscape of CIL for AA through the taxonomy of CIL approaches defined in this work.
\subsection{Class-Incremental Learning for AA}\label{cildef}

\begin{figure*}[t!]
  \centering
  \caption{Illustration of CIL for AA. The model first receives a set of authors along with multiple samples for training and evaluates its performance based on these authors. In subsequent sessions, the model updates itself only with new authors, different from those in previous sessions. Evaluation is performed using a combined set of all authors encountered thus far, including those introduced in the current session. Different colors correspond to different authors and their documents.}
  \includegraphics[width=\linewidth, height=3.5cm]{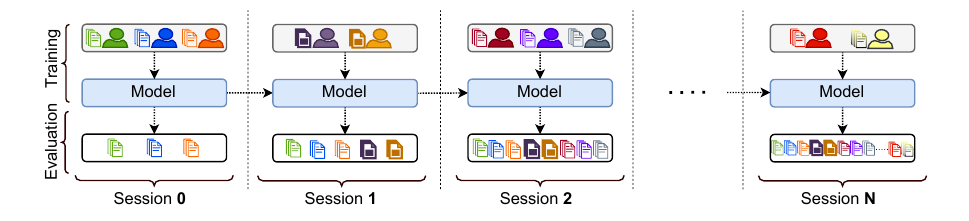}
  \label{CIL}
\end{figure*}

Class-Incremental learning (CIL) entails multiple sessions designed to simulate an ongoing learning process. \autoref{CIL} provides an overview of the steps involved in CIL. Typically each session is denoted as $S_n$. The CIL process begins with an initial training phase $S_0$, using dataset $D_0$ where:
\small{
\[D_0 = \{(a_{0i}, \{d_{0i,1}, d_{0i,2}, \ldots\}), \ldots, (a_{0m}, \{d_{0m,1}, d_{0m,2}, \ldots\})\}\]
}
Here, $a_{0i}$ represents author $i$ in the initial session, and $d_{0i,j}$ represents his $j$-th document. In subsequent sessions $S_t$ (where $t = 1, 2, ..., n$) the model encounters a variable number of new authors $k_t$:
\small{
\[D_t = \{(a_{t1}, \{d_{t1,1}, d_{t1,2}, \ldots\}), \ldots, (a_{tk_t}, \{d_{tk_t,1}, d_{tk_t,2}, \ldots\})\}\]
}
Where, there is no overlap of authors between sessions, ensuring $A_0 \cap A_t = \emptyset$ for all $t$, where $A_0$ and $A_t$ represent the sets of authors in $S_0$ and $S_t$, respectively. During these updates, the model leverages information solely from the current session $S_t$. Finally, the model's performance is evaluated across all previously encountered authors, including those introduced in the current session $S_t$.

\subsection{Class-Incremental Learning Landscape in AA}
Recently, CIL has garnered significant attention across various fields, including image classification, face recognition, and vehicle detection~\cite{jodelet2023class, hu2023dense, li2022few}. Its popularity is largely driven by the rise of large-scale models like GPT-2, where retraining the entire model for new classes is impractical. Furthermore, CIL is advantageous in situations where data privacy is crucial, as it enables models that do not retain training data~\cite{chamikara2018efficient}. Despite numerous efforts to categorize existing CIL approaches, a unified taxonomy remains elusive~\cite{masana2022class, zhou2023deep, mai2022online, tian2024survey}. This difficulty arises from the overlap between various techniques, making distinct categorization challenging. Therefore, to the best of our knowledge, for the first time within this work, we offered a high-level taxonomy of existing CIL approaches based on their level of ease in integrating with AA.

CIL primarily addresses the "stability-plasticity" dilemma~\cite{mermillod2013stability}, which involves balancing the retention of acquired knowledge (stability) with the ability to learn new tasks (plasticity). Extreme plasticity can lead to catastrophic forgetting~\cite{french1999catastrophic, mccloskey1989catastrophic}, where the model exclusively focuses on new data, resulting in the loss of previously learned information. Consequently, various approaches strive to achieve the optimal balance from different perspectives. 

In the following, we will provide a brief overview of the CIL landscape in AA w.r.t taxonomy.\newline


\subsubsection{Replay Method } This method involves replaying past data to help the model preserve the knowledge of previously learned classes while learning new ones in subsequent sessions. These sessions may use data from previous sessions or synthetically generated data through models such as Generative Adversarial Networks (GANs) or low-dimensional transformed representations~\cite{xiang2019incremental, shin2017continual}. Exemplar sampling strategies within replay methods vary widely. \textit{Random sampling}, despite its computational efficiency, has demonstrated effectiveness~\cite{chaudhry2018riemannian}. As research progressed, more sophisticated sampling strategies were introduced, which focused on selecting exemplars that are most informative for the learning process. For instance, choosing hard samples near decision boundaries or those with higher entropy derived from softmax outputs provided more value in retaining past knowledge and aiding the learning of new classes~\cite{chaudhry2018riemannian}. To further optimize memory usage, two primary approaches were proposed: \textit{fixed memory} and \textit{dynamic memory}. Fixed memory sizes offered simplicity but came with the overhead of managing the addition and removal of exemplars, as well as associated operational costs. On the other hand, dynamic memory allowed for the seamless addition of new examples, growing linearly with new data but requiring more storage space~\cite{chaudhry2018riemannian}.\newline

\subsubsection{Sequential Adaptation Strategy} This strategy is based on the concept of sequential learning, where it is assumed that the learner in session \( t-1 \) performed well in the previous classes and can transfer, adjust, or correct the learner in session \( t \). To achieve this, numerous approaches have been proposed, which can be further categorized into (I) \textit{parameter regularization}, (II) \textit{data regularization}, and (III) \textit{knowledge distillation}.\newline

\noindent\textbf{\textit{(I) Parameter Regularization.}} This concept relies on two assumptions: (1) not all model parameters contribute equally to classification, and (2) parameters are independent. The algorithm determines the importance of each parameter in session \( t-1 \) for classifying each author, copies and freezes learner parameters, and then updates the model with new authors in session $t$. When new authors are introduced to the model, it updates its parameters, by comparing current parameters to the frozen parameters from the previous session. Moreover, the second assumption of parameter independence ensures computational feasibility, given the typically large number of model parameters. To implement this, the model is penalized for significant changes in important parameters by adding a regularization term to the main loss $L_{\text{reg}} = \frac{1}{2} \sum_{i=1}^{|\theta^{t-1}|} \Omega_i (\theta^{t-1}_i - \theta^t_i)^2$. Here, $\theta^t_i$ is the current weight, $\theta^{t-1}_i$ is  the weight from session $t-1$, $|\theta^{t-1}|$ is the number of weights, and $\Omega_i$ represents the importance of the $i$th parameter. Different methods have been proposed to determine $\Omega$~\cite{kirkpatrick2017overcoming, liu2018rotate}.\newline

\noindent\textbf{\textit{(II) Data Regularization.}} Assuming that a learner in session  \( t-1 \) correctly classified the authors with access to their complete documents. Transferring this learner and some documents to session \( t \) can help the new learner adjust its behavior. This allows it to continue learning while maintaining performance similar to the previous frozen model. For instance,~\cite{lopez2017gradient} introduced the following regularization term into the main loss function:
{
\begin{equation}
\small{
\label{datareq}
\sum_{(x_{j},y_{j}) \in D_{0}^{t-1}} \ell(f_{\theta}(x_{j}),y_{j}) \leq \sum_{(x_{j},y_{j}) \in D_{0}^{t-1}} \ell(f_{\theta}^{t-1}(x_{j}),y_{j})
}
\end{equation}
}
Here, $f_{\theta}$ denotes the model with parameters $\theta$, and $f_{\theta}^{t-1}$ represents the model after training the previous session $S_{t-1}$. $D_{t}$ is the data from the current session, while $D_{0}^{t-1}$ refers to an exemplar set from sessions $D_{0}$ through $D_{t-1}$.  The loss term in~\autoref{datareq} encourages that the current model updates with the restriction that the loss on the transferred exemplars does not exceed the loss of the previous model. \newline

\noindent\textbf{\textit{(III) Knowledge Distillation. }} The knowledge distillation involves transferring knowledge from a larger, more complex model (the teacher) to a smaller, simpler model (the student)~\cite{hinton2015distilling}. In CIL, the previous model  \( f^{t-1} \) acts as the teacher, and the current model \( f^t \) is the student. The student model can use logit distillation to align its output probabilities with those of the teacher~\cite{rebuffi2017icarl, li2017learning}. Alternatively, feature distillation aims to replicate similar feature representations~\cite{hou2019learning, lu2022augmented}. In general, the primary advantage of Sequential Adaptation strategies lies in their ease of integration with existing models, particularly for AA models that commonly rely on pre-trained models~\cite{hicke2023t5, fabien2020bertaa}. However, one of their significant disadvantages can be their resource-intensive nature in terms of computation or memory usage (due to the necessity of maintaining two models or important matrix parameters) and the requirement to retrain the entire model, making them challenging for modern large-scale models. \newline

\subsubsection{Decoupled Learning} It is based on the idea that pre-trained models effectively extract general features from input samples. In this approach, models are divided into two components: \textit{the feature extractor}, which is a pre-trained model focusing solely on feature extraction, and \textit{the classifier}, which handles class discrimination~\cite{belouadah2018deesil, hayes2020lifelong}. The feature extractor is frozen after the initial training session to maintain its ability to extract general features and prevent degradation from over-finetuning. Meanwhile, the classifier is updated continuously in subsequent sessions. This method is particularly useful in scenarios with imbalanced datasets due to limited samples from new authors~\cite{li2022few}.\newline

\noindent Finally, to explore further details regarding the aforementioned approaches, please refer to the following works~\cite{jodelet2023class, hu2023dense, li2022few}.

\section{Related Work}
Several studies have endeavored to address AA in resembling real-world scenarios, one such endeavor is the open-set AA, which addresses scenarios where the true author of a text may not be among the set of candidate authors~\cite{badirli2019open, rahgouy2019cross}. This can reflect real-world applications, such as detecting malicious apps, where all malicious styles are considered potential candidates, and the unknown sample may be classified as a benign app. While open-set AA partially addresses real-world applications by accounting for new authors and grouping them into a single broad class, it is not entirely adequate for the nature of AA due to the need for further classification of new authors beyond what open-set AA offers. 

The work of Justin et al.~\cite{Justin20} is among the first works to adapt open-set AA for incremental learning. Given a dataset $D$, a subset of authors, k-seed, is used to train the model. During testing, a mixture of the k-seed known authors and k-new authors (authors in $D$ who were not in k-seed) is introduced to the model. The model is required to perform open-set classification and cluster the unknown samples into one or two clusters. The model is then retrained with both old and new unknown clustering samples, assigning unique labels to them. This process repeats until the model reaches a certain error rate. Although this approach presents a valuable method for incremental learning, it has limitations in real-world applications: (I) the model could be beneficial for tasks like malicious software detection, where detecting new types or styles is more critical than specific individual label classes but in most other AA problems cases, individual new class labels are known, and the model should only effectively differentiate between them. (II) The approach involves retraining the entire model, which is impractical for more advanced and large models such as BERT~\cite{devlin2018bert}. 

Moreover, the work by Luyckx et al.~\cite{luyckx2008authorship} examined the impact of having a limited number of documents per author on performance, concluding that lazy learners (e.g., KNN) outperform eager models (e.g., SVM) in such scenarios. Another study by \cite{luyckx2008authorship} investigated the impact of the number of authors on AA learners, revealing that high-performance results in the AA literature can be misleading due to the use of a small number of authors. By introducing a dataset containing 145 authors, they demonstrated a significant performance drop. Building upon this, we explore the effect of varying datasets within the CIL framework, including those with a larger number of authors—such as a dataset with 1,000 authors—and datasets that vary in the number of documents per author, ranging from a limited and small set as 10 documents to a more extensive set of over 1,000 documents per author. 

\section{Case Study}
In this work, we empirically validate the CIL paradigm in the context of AA upon several widely well-known AA datasets from multiple domains such as blog posts, news, reviews, and academic papers to ensure the findings and comprehensiveness of our study. Our empirical study covers a variety of scenarios from a limited number of authors to large-scale authors, and a limited number of documents per author to a substantial number of documents per author, to extensively verify the CIL paradigm use case, challenges, and future directions. In this section, we provide an overview of AA datasets, the process of converting them to the CIL paradigm, and final empirical validation sets as a case study of this work.

\subsection{Datasets}
The empirical validation of this work focuses on well-known AA datasets, which we will briefly discuss in this section and examples per dataset is provided at GitHub repository\footnote{\href{https://github.com/MostafaRahgouy/AA-CIL/tree/main/examples}{https://github.com/MostafaRahgouy/AA-CIL/tree/main/examples}}.

\begin{enumerate}
\item\textbf{IMDb62. }The first dataset of this study is, IMDb~\cite{seroussi2014authorship}, a prominent AA dataset containing movie reviews. The streaming nature of this dataset makes it well-suited for the CIL paradigm study. We maintained IMDb62, a subset of this dataset that contains 62 prolific users with 1,000 reviews per user, which contributes to overall 62,000 movie reviews~\cite{fabien2020bertaa,ai2022whodunit}.

\item\textbf{CCAT50. }The CCAT50~\cite{lewis2004rcv1} is a subset of the Reuters Corpus Volume I (RCV1), featuring news from the top 50 authors based on article volume. It includes 100 news per author, with a total of 5,000 news, all labeled with CCAT (corporate/industrial) subtopics to minimize topic influence in AA. 

\item\textbf{Blog50 and Blog1000. }The Blog corpus~\cite{schler2006effects} comprises relatively short blog posts collected from Blogger.com, featuring 19,000 authors and a total of 680,000 posts that subsets of such dataset extensively used in AA studies~\cite{fabien2020bertaa, ai2022whodunit}. For this study, we preprocessed the dataset and created two subsets named Blog50 and Blog1000. Blog50 comprises 50 authors, with each author having between 706 and 3,283 posts. On the other hand, to assess CIL in AA for a large-scale author set, we used Blog1000 subset, which consists of 1,000 authors, each with 100 posts. 
\item\textbf{arXiv100. }The arXiv \cite{arxiv23}, is a scholarly publication domain in the field of computer science and machine learning (abstracts of the research papers). The dataset consists of authors with at least 10 papers and authors with a maximum of 32 papers, resulting in a total of 1,469 documents from 100 authors. 
\end{enumerate}

\subsection{Building CIL Datasets} \label{section:builddataset}
For converting this study dataset into a CIL framework we developed a \texttt{\textsc{BuildCILData}} algorithm~\ref{algo}.  The algorithm takes a dataset (e.g., Blog50) with a list of session ratios and partitions‌ the dataset into multiple sessions for CIL. As an example, a dataset with a session ratio of [0.5, 0.5] (two sessions) partitions the dataset into two sessions, each containing half of the authors with all their documents. The algorithm begins by grouping all documents for each author. Next, per session, authors randomly selected according to the specified ratio by distributing author documents to 60\% for training, 20\% for validation, and 20\% remaining for testing within the session. This process is repeated for all sessions.  

Our CIL empirical validation consists of two scenarios. The first scenario consists of a \textbf{6-session} setup, which is built from an initial session with a ratio of 0.5 followed by 5 sessions each with a ratio of 0.1. The second scenario is called a \textbf{10-session} setup, which consists of 10 sessions and each session uses a ratio of 0.1. The first scenario is designed for pre-training model assessments, where a significant portion of data is reserved for the initial session. The second scenario, provides a different incremental learning setting, with smaller portions of data introduced in each session. Overall, both scenarios were created for a comprehensive assessment of the CIL paradigm.

\begin{algorithm}
\small
\label{algo}
\caption{\small \textsc{BuildCILData}: \textbf{iid} stands for \textit{Incremental ID} and \textbf{SESSIONS} is a list indicating the percentage of authors to include in each session. }
\begin{algorithmic}[1]
    \REQUIRE DATA, SESSIONS
    \STATE authors $\gets$ \{\}

    \FOR{\textbf{each} (author\_id, author\_content) \textbf{in} DATA}
        \IF{author\_id \textbf{not in} authors}
            \STATE authors[author\_id] $\gets$ []
        \ENDIF
        \STATE authors[author\_id]\textbf{.insert}(author\_content)
    \ENDFOR
    \STATE authors $\gets$ \textbf{shuffle}(authors)
    \STATE \textsc{CILData} $\gets$ [],  iid $\gets$ 0
    \FOR{\textbf{each} session \textbf{in} SESSIONS}
        \STATE session\_data $\gets$ authors[:session]
        \STATE authors $\gets$ authors[session:]
        \STATE train $\gets$ [], val $\gets$ [], test $\gets$ []
        \FOR{\textbf{each} (author\_id, contents) \textbf{in} session\_data}
            \STATE contents $\gets$ \textbf{shuffle}(contents)
            
            \STATE train\_contents $\gets$ contents*0.6
            \STATE train\textbf{.insert}(\{iid, author\_id, train\_contents\}) 
            
            \STATE val\_contents $\gets$ contents*0.2
            \STATE val\textbf{.insert}(\{iid, author\_id, val\_contents\}) 
            
            \STATE test\_contents $\gets$ contents*0.2
            \STATE test\textbf{.insert}(\{iid, author\_id, test\_contents\}) 
            \STATE iid $\gets$ iid + 1
        \ENDFOR
        \STATE \textsc{CILData}\textbf{.insert}([train, val, test])
    \ENDFOR
    
    \STATE \textbf{Return} \textsc{CILData} 
\end{algorithmic}
\end{algorithm}

\subsection{Empirical Validation Sets}
\begin{figure}[t]
  \centering
  \caption{Empirical validation set categories for 5 datasets.}
  \includegraphics[width=3cm, height=3cm]{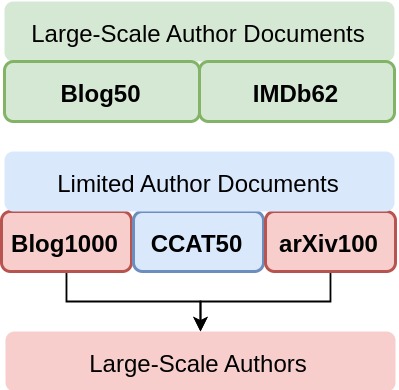}
  \label{CILDatasetCategory}
\end{figure}

Our obtained empirical validation set w.r.t the study dataset and \texttt{\textsc{BuildCILData}} algorithm, is categorized into three groups as presented in~\autoref{CILDatasetCategory}, \texttt{Large-Scale Author Documents}, \texttt{Limited Author Documents}, and \texttt{Large-Scale Authors}. The statistics of the datasets for the 6-session setup are presented in~\autoref{6_sessions_data}, and for the 10-session setup presented in~\autoref{10_sessions_data}. According to statistics tables, we can see that \texttt{Large-Scale Author Documents} datasets Blog50 and IMDb62 contain large amounts of documents at each session per author, on the other hand, \texttt{Limited Author Documents} datasets Blog1000, CCAT50, and arXiv100 has lower amount of documents at each session per author. Moreover, we considered Blog1000 and arXiv100 datasets in this study as a \texttt{Large-Scale Authors} due to the high amount of authors presented at each session. The dataset presented in this study is from four different domains such as news (CCAT50), blog posts (Blog50, Blog1000), reviews (IMDb62), and publications (arXiv100). 

One specific design consideration is having 2 different scenarios with sessions where for 6-session setup, we assign 50\% of the authors (within their documents) to initial session $s0$ and in rest of the remaining 50\% split into the remaining 5 sessions, this can be observed in~\autoref{6_sessions_data} where we have large authors and documents for $s0$ than the other sessions. Nevertheless, for 10-session setup, we assigned 10\% of authors per session starting from initial session $s0$ to the final session $s9$. This supports two real-world scenarios where in one case we may have a database of authors but keep adding new authors over time, in another case we want to build a system that is designed for specific settings such as building an AA application for university student homework. In summary, considerations of this study support completeness of our studies from different domains, scenarios, and perspectives such as different levels of granularity w.r.t number of documents and authors.

\begin{table*}[h!]
\centering
\tiny
\renewcommand{\arraystretch}{.98}
\caption{Dataset splits per session (6-session setup). 'S' column shows splits, for TR (Train), V (Validation), and TS (Test) at each session.}
\begin{tabular}{|p{1.1cm}|p{0.3cm}|p{0.25cm}|p{0.6cm}|p{0.25cm}|p{0.55cm}|p{0.25cm}|p{0.55cm}|p{0.25cm}|p{0.55cm}|p{0.25cm}|p{0.55cm}|p{0.4cm}|p{0.55cm}|}
\hline
\multirow{2}{*}{Dataset} & \multirow{2}{*}{S} & \multicolumn{12}{c|}{Sessions} \\ \cline{3-14}
 &  & \multicolumn{2}{c|}{s0} & \multicolumn{2}{c|}{s1} & \multicolumn{2}{c|}{s2} & \multicolumn{2}{c|}{s3} & \multicolumn{2}{c|}{s4} & \multicolumn{2}{c|}{s5} \\ \cline{3-14}
&  & A & D & A & D & A & D & A & D & A & D & A & D \\ \cline{1-14} 
\multirow{3}{*}{Blog50} & TR & 25 & 16496 & 5 & 2729 & 5 & 2636 & 5 & 2877 & 5 & 3772 & 5 & 3627 \\ 
 & V & 25 & 5496 & 5 & 911 & 5 & 879 & 5 & 958 & 5 & 1258 & 5 & 1211 \\ 
  & TS & 25 & 5501 & 30 & 6410 & 35 & 7288 & 40 & 8,247 & 45 & 9505 & 50 & 10713 \\  
\hline
\multirow{3}{*}{IMDB62} & TR & 31 & 18,600 & 6 & 3,600 & 6 & 3600 & 6 & 3600 & 6 & 3,600 & 7 & 4200 \\ 
 & V & 31 & 6200 & 6 & 1200 & 6 & 1200 & 6 & 1200 & 6 & 1200 & 7 & 1400 \\ 
  & TS & 31 & 6178 & 37 & 7376 & 43 & 8576 & 49 & 9772 & 55 & 10970 & 62 & 12370 \\  
\hline
\multirow{3}{*}{CCAT50} & TR & 25 & 1,500 & 5 & 300 & 5 & 300 & 5 & 300 & 5 & 300 & 5 & 300 \\ 
 & V & 25 & 500 & 5 & 100 & 5 & 100 & 5 & 100 & 5 & 100 & 5 & 100 \\ 
  & TS & 25 & 500 & 30 & 600 & 35 & 700 & 40 & 800 & 45 & 900 & 50 & 1000 \\  
\hline
\multirow{3}{*}{Arxiv100} & TR & 50& 438&  10&  72&  10&  73&  10&  76&  10& 85 & 10 & 83 \\ 
 & V & 50 & 161 & 10 & 28 & 10 & 28 & 10 & 27 & 10 & 33 & 10 & 30 \\ 
  & TS & 50 & 173 & 60 & 204 & 70 & 236 & 80 & 268 &  90 & 303 & 100 & 335 \\  
\hline
\multirow{3}{*}{Blog1000} & TR & 500&  30000&  100&  6000&  100&  6,000&  100&  6000&  100&  6000&  100& 6000 \\ 
 & V &  500&  10000&  100&  2000&  100&  2000&  100&  2000&  100&  2000&  100& 2000 \\ 
  & TS &  500&  10000&  600&  12000&  700&  14000&  800&  16000&  900&  18000&  1000& 20000 \\  
\hline
\end{tabular}
\label{6_sessions_data}
\end{table*}

\begin{table*}[h!]
\centering
\tiny
\renewcommand{\arraystretch}{0.98}
\caption{Dataset splits per session (10-session setup). 'S' column shows splits, for TR (Train), V (Validation), and TS (Test) at each session.}
\begin{tabular}{|p{1.1cm}|p{0.3cm}|p{0.28cm}|p{0.45cm}|p{0.28cm}|p{0.46cm}|p{0.28cm}|p{0.46cm}|p{0.28cm}|p{0.46cm}|p{0.28cm}|p{0.55cm}|p{0.28cm}|p{0.56cm}|p{0.29cm}|p{0.56cm}|p{0.28cm}|p{0.56cm}|p{0.28cm}|p{0.56cm}|p{0.4cm}|p{0.56cm}|}
\hline
\multirow{2}{*}{Dataset} & \multirow{2}{*}{S} & \multicolumn{20}{c|}{Sessions} \\ \cline{3-22}
 &  & \multicolumn{2}{c|}{s0} & \multicolumn{2}{c|}{s1} & \multicolumn{2}{c|}{s2} & \multicolumn{2}{c|}{s3} & \multicolumn{2}{c|}{s4} & \multicolumn{2}{c|}{s5} & \multicolumn{2}{c|}{s6} & \multicolumn{2}{c|}{s7} & \multicolumn{2}{c|}{s8} & \multicolumn{2}{c|}{s9} \\ \cline{3-22}
&  & A & D & A & D & A & D & A & D & A & D & A & D & A & D & A & D & A & D & A & D \\ \cline{1-22} 
\multirow{3}{*}{Blog50} & TR & 5 & 3398 & 5 & 2985 & 5 & 3323 & 5 & 4309 & 5 & 2481 & 5 & 2729 & 5 & 2636 & 5 & 2877 & 5 & 3772 & 5 & 3627 \\ 
 & V & 5 & 1132 & 5 & 995 & 5 & 1107 & 5 & 1436 & 5 & 826 & 5 & 911 & 5 & 879 & 5 & 958 & 5 & 1258 & 5 & 1211 \\ 
  & TS & 5 & 1133 & 10 & 2128 & 15 & 3236 & 20 & 4673 & 25 & 5501 & 30 & 6410 & 35 & 7288 & 40 & 8247 & 45 & 9505 & 50 & 10713 \\  
\hline
\multirow{3}{*}{IMDB62} & TR & 6 & 3600 & 6 & 3600 & 6 & 3600 & 6 & 3600 & 6 & 3600 & 6 & 3600 & 6 & 3600 & 6 & 3600 & 6 & 3600 & 8 & 4800 \\ 
 & V & 6 & 1200 & 6 & 1200 & 6 & 1200 & 6 & 1200 & 6 & 1200 & 6 & 1200 & 6 & 1200 & 6 & 1200 & 6 & 1200 & 8 & 1600 \\ 
  & TS & 6 & 1179 & 12 & 2379 & 18 & 3579 & 24 & 4778 & 30 & 5978 & 36 & 7176 & 42 & 8376 & 48 & 9572 & 54 & 10770 & 62 & 12370 \\  
\hline
\multirow{3}{*}{CCAT50} & TR & 5 & 300 & 5 & 300 & 5 & 300 & 5 & 300 & 5 & 300 & 5 & 300 & 5 & 300 & 5 & 300 & 5 & 300 & 5 & 300 \\ 
 & V & 5 & 100 & 5 & 100 & 5 & 100 & 5 & 100 & 5 & 100 & 5 & 100 & 5 & 100 & 5 & 100 & 5 & 100 & 5 & 100 \\ 
  & TS & 5 & 100 & 10 & 200 & 15 & 300 & 20 & 400 & 25 & 500 & 30 & 600 & 35 & 700 & 40 & 800 & 45 & 900 & 50 & 1000 \\  
\hline
\multirow{3}{*}{Arxiv100} & TR & 10& 88& 10& 87& 10& 69& 10& 95& 10& 99& 10& 72& 10& 73& 10& 76& 10& 85& 10& 83 \\ 
 & V & 10& 32& 10& 32& 10& 26& 10& 36& 10& 35& 10& 28& 10& 28& 10& 27& 10& 33& 10& 30  \\ 
  & TS & 10& 35& 20& 69& 30& 97& 40& 134& 50& 173& 60& 204& 70& 236& 80& 268& 90& 303& 100& 335 \\  
\hline
\multirow{3}{*}{Blog1000} & TR & 100& 6000& 100& 6000& 100&  6000& 100& 6000& 100& 6000& 100& 6000& 100& 6000& 100& 6000& 100& 6000& 100& 6000\\ 
 & V & 100& 2000& 100& 2000& 100& 2000& 100& 2000& 100& 2000& 100& 2000& 100& 2000& 100& 2000& 100& 2000& 100& 2000  \\ 
  & TS & 100& 2000& 200& 4000& 300& 6000& 400& 8000& 500& 10000& 600& 12000& 700& 14000& 800& 16000& 900& 18000& 1000& 20000 \\  
\hline
\end{tabular}
\label{10_sessions_data}
\end{table*}

\section{Methodology: Authorship Attribution Incremental Learners}
As observed in the CIL steps within~\autoref{CIL} and defined the CIL in the~\autoref{cildef}, the \texttt{Model} is the main learner which incorporates new authors in training at each session and makes an evaluation based on the whole evaluation data including new authors' documents. In this regard, we developed an AA incremental learner model with two main components: \textit{a pre-trained feature extractor} and a set of \textit{classifiers or heads for author classification}. For new authors (new sessions), we add a new head with weights corresponding to the number of new authors. We considered a pre-trained BERT-Base transformer model as a backbone of the proposed framework, leveraging its [CLS] token for the overall representation of an author's document. For classifiers, linear transformations were added to the top of the BERT model. However, training the feature extractor from scratch in this framework is a possibility we determined to use the pre-trained BERT checkpoints due to their powerful generalization capabilities in enhancing the AA system. Using the feature extractor and its multi-heads, we adapted several variants of the proposed AA incremental learner frameworks to address AA incremental learning. The details of each variant are as follows:

\subsection{Finetuning-based Frameworks} 
Finetuning is a straightforward approach to class incremental learning where the model, including the feature extractor and heads, is finetuned to incorporate new authors in subsequent sessions. This method uses cross-entropy loss for training and serves as a baseline to highlight catastrophic forgetting, where the model forgets previously learned information after introducing new sessions. We proposed two variants of this approach as follows: \textbf{FT}: The entire model, including both the feature extractor and all heads, is fine-tuned for new sessions; \textbf{FT+}: Only the backbone and the heads for the current session are updated in non-initial sessions, while previously learned heads remain unchanged, helping mitigate catastrophic forgetting.

\subsection{Decoupled-based Frameworks}

The decoupled-based paradigm restricts updates to the feature extractor after the initial session, leveraging a pre-trained model fixed after initial adaptation to handle the AA task. This approach assumes that the feature extractor, having processed a significant portion of data initially (typically 50\% in the initial session), does not require further learning and can provide stable representations while classifiers/heads handle further discrimination. We proposed two variants of this approach as follows: \textbf{FZ}: The BERT model acts as a fixed feature extractor after the initial session, with updates applied to all heads in subsequent sessions; \textbf{FZ+}: The BERT model remains a fixed feature extractor post-initial session, with updates limited to the heads relevant to the current session, akin to the FT+ approach.

\subsection{Knowledge Distillation-based Framework}
For the knowledge distillation approach, we considered \textbf{Learning Without Forgetting (LWF)} framework introduced by~\cite{li2017learning}, where LWF incorporate the following loss:
\begin{equation}
\small{
L = \mathcal{L}(f_t(x), y) + \sum_{k=1}^{K_{t-1}} -S_k^{t-1}(f_{t-1}(x)) \log S_k^{t}(f_t(x))
}
\end{equation}
where \( \mathcal{L}(f_t(x), y) \) denotes the standard cross-entropy loss between the predictions of the current model \( f_t \) and true labels \( y \). The second term in the loss formula represents the distillation loss, where \( S_k^{t-1}(f_{t-1}(x)) \) and \( S_k^{t}(f_t(x)) \) are the softmax probabilities of the $k$-th class outputted by models \( f_{t-1} \) and \( f_t \), respectively. This term aims to align the predicted probabilities of the previous model \( f_{t-1} \) with those of the current model \( f_t \), thereby transferring knowledge from the previous task \( t-1 \) to the current task \( t \). By minimizing this loss, LWF ensures that the model maintains performance on previous tasks while learning new ones sequentially, contributing to effective continual learning without forgetting.

\subsection{Replay-based Frameworks}
Replay models use exemplars from previous sessions to mitigate catastrophic forgetting and improve model performance. The models are fintuned using these exemplars, with the number of exemplars influencing the training process and effectiveness. We applied two variants of replay-based frameworks as follows:  \textbf{FT-Ek}, finetunes the model using $k$ randomly selected exemplars per author.  \textbf{LWF-Ek}, that applies the LWF model using an exemplar set with $k$ exemplars per author, e.g. LWF-E2 which is uses two examples.

\subsection{Parameter Regularization-based Frameworks}
Weight regularization methods aim to mitigate catastrophic forgetting by preserving important parameters of the model. Elastic Weight Consolidation (EWC)~\cite{kirkpatrick2017overcoming} and Memory Aware Synapses (MAS)~\cite{aljundi2018memory} are two such techniques that achieve this by evaluating and maintaining the importance of model parameters. We proposed two variants of parameter regularization-based frameworks as follows: \textbf{EWC:} Computes weight importance using a diagonal approximation of the Fisher Information Matrix and regularizes weights based on their estimated importance to previous tasks; \textbf{MAS}: Accumulates an importance measure for each parameter based on its sensitivity to changes in the predicted output function and regularizes weights by considering their impact on the model’s output stability.

\section{Results}

\subsection{Evaluation Metrics and Training Details}

\subsubsection{Metrics} To evaluate the models' performance, we used two common metrics: \textbf{Performance Drop rate (PD)} and \textbf{Average Accuracy (AvgA)}. The PD rate measures the accuracy difference between the final incremental and base sessions (lower PD is preferred) and AvgA gives the average accuracy across all sessions. \label{sec-eval}\newline

\subsubsection{Training Details} For all experiments, we utilized the pre-trained BERT-base-uncased model as our backbone network. We selected a batch size of 32 and trained each session for 5 epochs. To ensure optimal performance, validation sets were employed to save the best model for each session. The AdamW optimizer, with its default parameters, was used for training. All experiments were conducted on a single NVIDIA A100 GPU with 40 GB of memory. To guarantee the reproducibility of the results, we applied seeding. Furthermore, for the other aforementioned models, we retained all their proposed parameters as default.

\subsection{AA Incremental Learners Results}

\begin{table*}[t]
\centering
\renewcommand{\arraystretch}{1.4}
\scriptsize
\caption{Average accuracy and performance drop rate (PD) for models on \textbf{Blog50} and \textbf{IMDB62} datasets in \textbf{6-Session} Setup. The initial session average accuracy values of \textbf{83.86\%}, 
 and \textbf{98.37\%} respectively (initial session s0 results not depicted).}
\begin{tabular}{|c|*{14}{p{0.63cm}|}} \hline
\multirow{2}{*}{Model} & \multicolumn{6}{c|}{Blog50} & \multicolumn{6}{c|}{IMDb62}  \\ \cline{2-13} 
 &  s1 & s2 & s3 & s4 & s5 & PD $\downarrow$ & s1 & s2 & s3 & s4 & s5 & PD $\downarrow$ \\ \cline{2-13} \hline
 FT  & 30.9& 14.08& 11.92& 12.29& 11.02&                72.84&       34.18& 18.88& 12.28& 11.28& 11.27&  87.17        \\ \hline
 FT+  & 69.98& 61.16& 46.71& 29.88& 22.52&         61.34&       82.89& 61.91& 40.46& 29.12& 17.77& 80.6         \\ \hline
 \hline
 FZ  & 72.31& 64.67& 57.05& 47.89& 37.8&               46.06&       84.3& 75.82& 68.39& 59.8& 48.59&  49.78     \\   \hline
 FZ+  & 71.97& 63.3& 55.94& 48.53& 43.05&              40.81&       82.39& 70.86& 62.19& 55.4& 49.13&  49.24     \\ \hline \hline
 
 LWF  & \textbf{75.18}& 65.36& 57.35& 50.39& 43.11&             40.75&       \textbf{93.67}& 87.48& 77.97& 69.6& 55.34&  43.03  \\ \hline
 EWC  & 71.83& 65.63& 52.98& 41.79& 32.75&             51.11&       75.66& 63.27& 56.58& 54.44& 43.54& 54.83     \\ \hline
 MAS  & 70.19& 61.68& 54.55& 47.33& 42.25&             41.61&       87.12& 74.5& 65.63& 58.56& 52.15& 46.22       \\ \hline \hline
 
 FT\_E2  & 43.7& 38.76& 34.87& 34.71& 32.24&           51.62&       81.16& 74.03& 61.47& 60.57& 56.95&  41.42       \\ \hline
 LWF\_E2  & 54.71& 36.91& 29.59& 31.16& 30.11&         53.75&       87.35& 72.98& 59.09& 55.38& 47.74& 50.63       \\ \hline
 

 FT\_E10  & 62.22& 59.99& 54.7& 50.87& 48.17&  35.69    &     92.64 & 90.64& 86.15& 85.81 & 81.96&  16.41      \\ \hline
 LWF\_E10  & 62.5& 58.48& 49.79& 48.2& 46.58&  37.28   &       89.9& 87.48& 78.33& 77.89& 75.25& 23.12    \\ \hline
\hline
 FT\_E20  & 69.69& 67.36& 58.68& 61.53& 58.74&  \textbf{25.12}           &       95.50& 93.33& 91.82& 90.46& 88.23&  \textbf{10.14}      \\ \hline
 LWF\_E20  & 69.03& 64.81& 57.97& 56.37& 57.48&   \textbf{26.38}                  &       93.91& 90.83& 87.68& 85.59& 83.04& \textbf{15.33}       \\ \hline
\end{tabular}
\label{CIL-results-largeaa}
\end{table*}

\begin{table*}[t]
\centering
\renewcommand{\arraystretch}{1.4}
\scriptsize
\caption{Average accuracy and performance drop rate (PD) for models on \textbf{Blog1000}, \textbf{CCAT50}, and \textbf{ArXiv100} datasets in \textbf{6-session} setup. The initial session average accuracy values of \textbf{50.61\%}, \textbf{87.6\%}, and \textbf{80.92\%} respectively (initial session s0 results not depicted).}
\begin{tabular}{|c|*{20}{p{0.515cm}|}}
\hline
\multirow{2}{*}{Model} & \multicolumn{6}{c|}{Blog1000}  & \multicolumn{6}{c|}{CCAT50} & \multicolumn{6}{c|}{ArXiv100}\\ \cline{2-19} 
 &  s1 & s2 & s3 & s4 & s5 & PD $\downarrow$ & s1 & s2 & s3 & s4 & s5 & PD $\downarrow$ & s1 & s2 & s3 & s4 & s5 & PD $\downarrow$ \\ \cline{2-19} \hline
 FT        &   13.28& 8.95& 8.52& 7.84& 6.87&  43.74    &       28.99& 19.0& 14.62& 12.33& 10.6&  77&     27.93& 16.1& 14.93& 11.21& 8.66& 72.26 \\ \hline
 FT+       &   21.92& 12.78& 9.92& 9.12& 8.26&  42.35      &       60.33& 33.29& 40.25& 22.43& 23.3& 64.3 &        67.16& 50.84& 38.81& 23.43& 20.0& 60.92 \\ \hline
 \hline
 FZ        &   42.71& 33.35& 25.35& 19.93& 15.64& 34.97    &       73.33&  62.86& 55.0& 48.89& 44.1& 43.5 &     68.14& 59.31& 51.87& 46.2& 42.39& 38.53 \\ \hline
 FZ+       &   42.18& 36.15& 31.63& 28.12& 25.3& \textbf{25.31}     &       73.0& 62.57& 54.75& 48.67& 43.8& 43.8 &     68.63& 59.31& 52.23& 46.2& 41.79& 39.13 \\ \hline \hline
 
 LWF       &   \textbf{44.49}& 34.55& 24.06& 18.84& 16.91& 33.7     &      \textbf{78.0}& 68.57& 59.88& 48.55& 44.7& 42.9 &      70.59& 59.31& 45.9& 37.62& 30.75& 50.17 \\ \hline
 EWC       &   22.82& 18.32& 18.65& 16.64& 15.29& 35.32    &      49.17& 38.86& 35.88& 22.0& 26.2& 61.4  &     64.71& 51.27& 35.07& 23.76& 19.1& 61.82 \\  \hline
 MAS       &   42.24& 36.21& 31.66& 28.13& 25.3& \textbf{25.31}     &      72.5& 62.29& 55.75& 50.11& 46.0& 41.6 &     72.06& 59.31& 52.23& 46.53& 43.58& 37.34 \\ \hline \hline
 
 FT\_E2    &  32.32& 28.22& 25.91& 23.21& 20.86& 29.75     &       69.33& 61.29& 63.62& 55.33& 60.0& \textbf{27.6} &      73.53& 71.61& 67.91& 67.0& 62.09& \textbf{18.83} \\ \hline
 LWF\_E2   &  37.8& 28.93& 24.24& 23.01& 22.58& 28.03     &       70.83& 58.86& 59.13& 50.44& 54.5& 33.1 &      68.14& 66.53& 64.92& 64.36& 60.0& 20.92 \\ \hline
\end{tabular}
\label{CIL-results-limitedaa}
\end{table*}

\begin{figure*}[t]
  \caption{Averaged Accuracy of Models on Datasets for \textbf{10-session} setup.} 
  \centering
  \includegraphics[width=\textwidth, height=4cm]{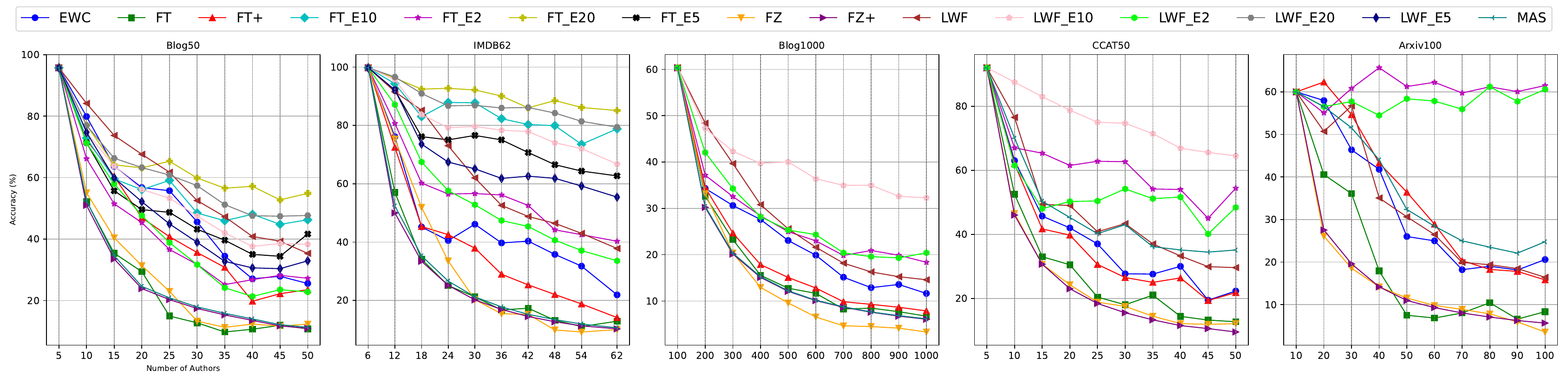}
  \label{inc-acc-10-sessions}
\end{figure*}

The \autoref{CIL-results-largeaa} summarize the results of our experiments for 6-session setup on the \texttt{Large-Scale Author Documents} datasets. Similarly, the \autoref{CIL-results-limitedaa} present results for a 6-session setup on the \texttt{Limited Author Document} and \texttt{Large-Scale Authors} datasets. Moreover,  experimental results for the 10-session setup are visualized on \autoref{inc-acc-10-sessions}. In the following sections, we will discuss CIL paradigm experimental results, beginning with the challenges posed by the datasets and will move forward with their compatibility with different models for sessions and explore the findings.

\subsubsection{Difficulty of the Datasets}
In our analysis, we evaluated dataset difficulty over a 6-session setup, focusing on the initial session where 50\% of the data was allocated. This large data portion typically yields higher model performance, setting a strong baseline for understanding dataset difficulty. As described in Section~\ref{section:builddataset}, this 6-session setup is designed for pre-training models (FT), with the initial session using a significant data portion. Subsequent sessions involve CIL based on this pre-trained model.  In the initial session, the FT model achieved accuracies of 98.37\% for IMDB62, 83.86\% for Blog50, 80.92\% for ArXiv100, 87.6\% for CCAT50, and 50.61\% for Blog1000. Thus, suggesting the datasets sorted by difficulty as Blog1000 (most difficult) $<$ ArXiv100 $<$ CCAT50 $<$ Blog50 $<$ IMDB62 (easiest). Challenges are generally greater when the number of authors increases and when there are fewer documents per author, as seen with the difficulties associated with \texttt{Large-Scale Authors} datasets (Blog1000 and ArXiv100).


\subsubsection{Severity of Catastrophic Forgetting}
With the CIL scenario, all previous and new classes are evaluated. FT is the most comment baseline that learns each session incrementally while not using any knowledge from previous sessions, while, FT+ is another variant that avoids slow forgetting. According to FT model results in a 6-session setup, it is apparent that IMDb62 with PD of 87.17\% experiences severe forgetting, next, CCAT50 with PD of 77\%,  is suffering from severe forgetting, ArXive and Blog50 with PD of $\approx 72.5\%$, and lastly, Blog1000 with PD of 43.74\% are showing high severity in compared to other models. However, the FT+ slows down the forgetting by an overall average of $\approx 9\%$, for datasets, but it still suffers from forgetting in comparison with other models. FT+ slows down the overall forgetting behavior of FT, however, in large-scale authors, it is a different story as we can see that with Blog1000 we only obtained a 1.38\% improvement in PD.  Moreover, analysis of the 10-session setup, for all datasets, the ~\autoref{inc-acc-10-sessions}, confirms that FT (green-colored) drops quickly compared to FT+ (red-colored). Recent works of~\cite{yu2020semanticdriftcompensationclassincremental} show that the cross-entropy loss might be responsible for high levels of catastrophic forgetting. They reported that less forgetting can be achieved by replacing the cross-entropy loss with a metric learning loss or by using an energy-based method.
\subsubsection{Freezing Feature Extractor After Initial Session}
The decoupled-based frameworks (FZ and FZ+) are designed to freeze the backbone model (feature extractor), allowing them to outperform the FT and FT+ models by only updating the heads relevant to the current session. Both models (FZ and FZ+), by retaining more information from previous sessions, show significant improvements across datasets, especially in the 6-session setup where it benefits from learning over 50\% of the data during the initial session. However, in the 10-session setups (orange and purple-colored), these frameworks did not surpass FT and FT+, and in Blog1000 performed poorly in comparison to all other models. This highlights a limitation of decoupled-based frameworks: while they excel in generalizing knowledge over multiple sessions due to the initial learning phase (as seen in the 6-session setup), they may lose effectiveness when there is a high number of sessions --e.g 10-session setup-- and are less information to learn at the initial session and new sessions, resulting in poorer performance compared to FT models in the 10-session setup. Additionally, in the 6-session setup, FZ+ showed minimal improvement ($\approx$1\%) over FZ and struggled with CCAT50, ArXiv100, and IMDb62, the same behavior was also seen in the 10-session setup.

\subsubsection{Performance Dynamics in Sequential Learning}
\begin{figure*}[t]
  \centering
    \caption{Confusion matrix of various models on Blog50 (upper row) and CCAT50 (lower row) after last incremental stage (\textbf{6-session} setup).}
  \includegraphics[width=\linewidth, height=6.5cm]{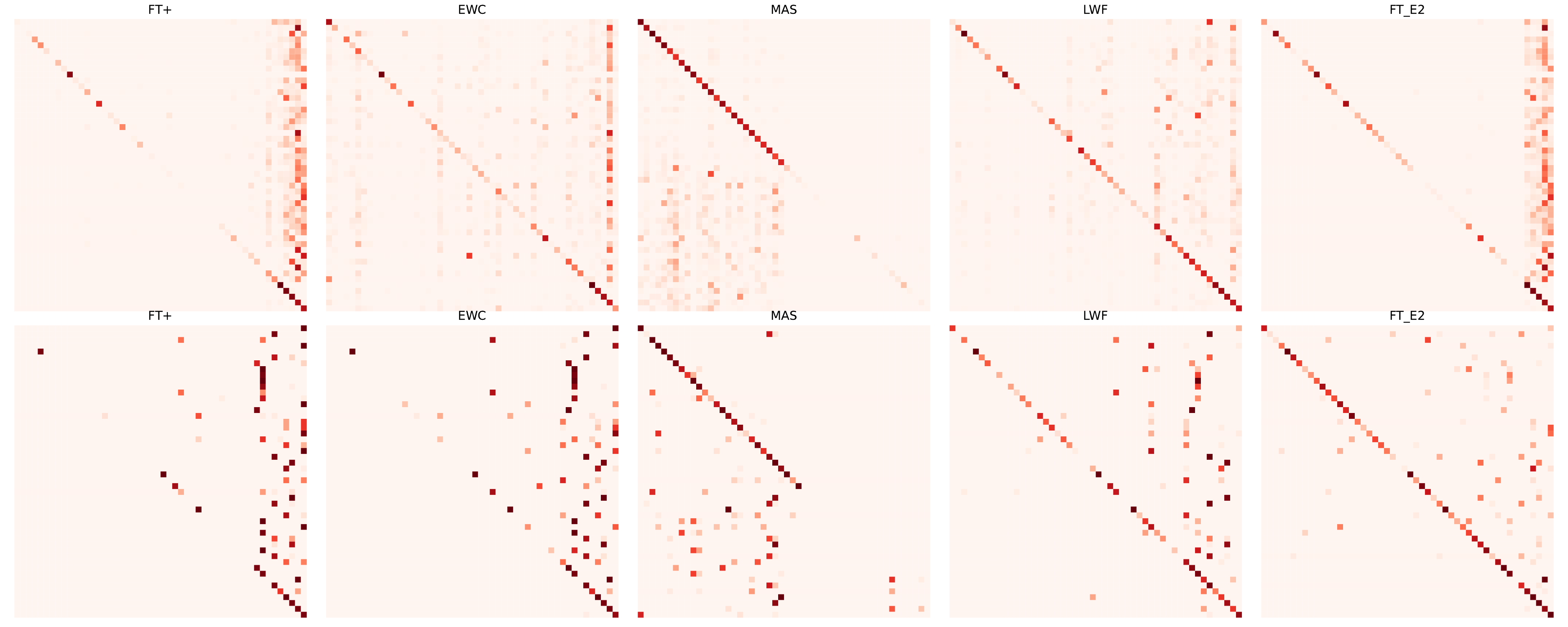}
  \label{conf-matrix}
\end{figure*}

The results for LWF (knowledge distillation), EWC (weight regularization), and MAS (parameter regularization) in 6-session setups highlight LWF's superiority in \texttt{Large-Scale Author Documents} datasets, with PDs of 40.75\% for Blog50 and 43.03\% for IMDb62. On the other hand, MAS demonstrates its strength in \texttt{Limited Author Documents} datasets, achieving PDs of 25.31\% for Blog1000, 41.60\% for CCAT50, and 37.34\% for ArXiv100 in the 6-session setups. The 10-session setup reveals similar behavior, with MAS (light green) and LWF (dark red) consistently showing their respective strengths. One possible reason why LWF is performing well with \texttt{Large-Scale Author Documents} dataset can be referred to as weight updates, where batch updates with larger documents per author don't lead to weight overestimating the importance, unlike what is observed with MAS. On the other hand, MAS performs better with \texttt{Limited Author Documents} datasets because it effectively captures the sensitivity of the learned function (reflected in the gradient's magnitude)~\cite{aljundi2018memory} when data is limited, resulting in better performance. In session $s1$ of the 6-session setup, LWF performed exceptionally well (values in bold in Tables \ref{CIL-results-largeaa} and \ref{CIL-results-limitedaa}), due to effectively retained knowledge, benefiting from preserving distributions with the help of the teacher model. However, as the model encountered less data in subsequent sessions, particularly with \texttt{Limited Author Documents}, it began to forget. 

Based on the confusion matrix as presented in \autoref{conf-matrix}, for the EWC, MAS, and LWF models on the \texttt{Limited Author Document} dataset CCAT50 and the \texttt{Large-Scale Author Documents} dataset Blog50 for the 6-session setup, we observed the following: 
\begin{itemize}
    \item \textbf{Blog50. }First, the MAS model begins to experience severe forgetting in the middle sessions, leading to a significant drop in performance compared to the early stages. Second, LWF shows fewer errors than EWC and is more successful in capturing new authors while retaining the knowledge of previous authors. However, both models struggle to distinguish between authors when introducing new authors. This is evident in the upper triangular matrix of LWF (showing fewer errors compared to EWC) and the errors in both the upper and lower triangular matrices for EWC (indicating a higher error rate).
    \item \textbf{CCAT50. }EWC displays significant catastrophic forgetting, as indicated by a PD rate of 61.4\%, similar to the behavior observed with FT+. LWF (with a PD rate of 42.9\%) and MAS (PD of 41.6\%) models perform similarly on the CCAT50 dataset. Despite MAS advantage compared to previous models in \texttt{Limited Author Document} datasets, it tends to over-memorize previous sessions, leading to difficulties in learning new information. 
\end{itemize}

\begin{figure}[t]
  \centering
    \caption{Results of PD rate using 2, 5, 10, and 20 examples for FT-EK and LWF-EK models using Blog50 and IMDb62 datasets at 6-session setup.}
  \includegraphics[width=\linewidth, height=4.3cm]{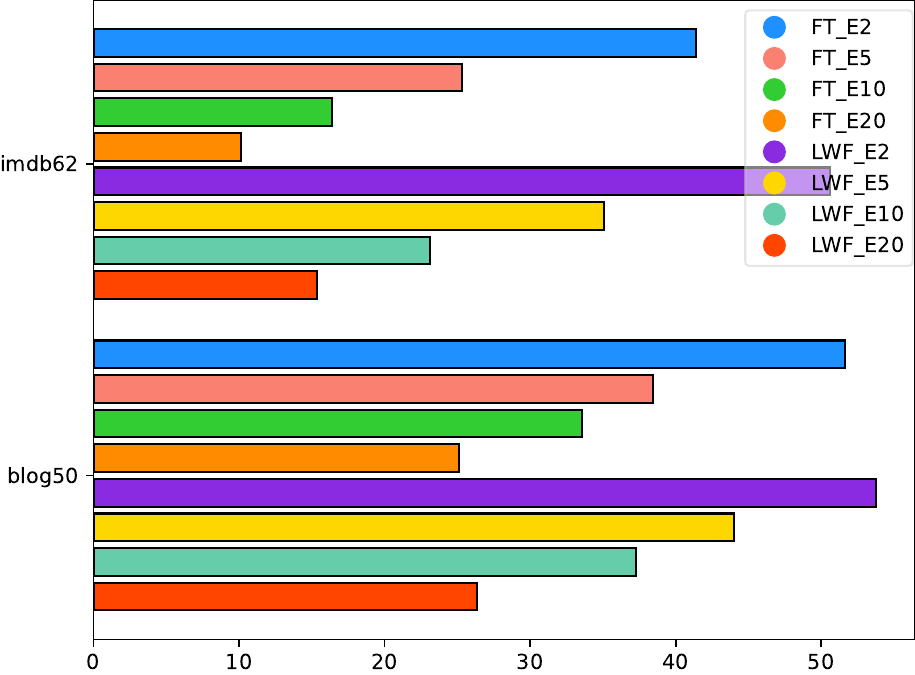}
  \label{pd-analysis}
\end{figure}

\subsubsection{Exemplar Rehearsal -- Replay with Random Examples}
In Exemplar Rehearsal methods, attempting to increase the number of examples in ArXiv100, where there are fewer documents per author, leads to a violation of CIL principles. This is because CIL aims to minimize reliance on previous session data, and adding more exemplars contradicts this goal, especially given the limited data available in ArXiv100. Because of this, we didn't extend experimentations with \texttt{Limited Author Documents}. 

\noindent\textbf{FT-based Exemplar Rehearsal. }The results from the 6-session and 10-session setups indicate that the FT-E2 model, which employs two randomly selected exemplars per author, significantly improved performance on the IMDB62 dataset, achieving a PD rate of 41.42\%, outperforming other frameworks -- in 6-session setup. This advantage was not observed with the Blog50 dataset, likely due to its higher difficulty. Larger exemplar sets, such as those in FT-E10 or FT-E20, showed marked improvements in \texttt{Large-Scale Author Documents} datasets including Blog50, where the volume of documents per author supports extensive exemplar rehearsal. However, in \texttt{Limited Author Documents} datasets, smaller exemplar sets proved more effective due to fewer documents per author, as observed with CCAT50 (PD rate of 27.6\%) and ArXiv100 (PD rate of 18.83\%). For datasets like ArXiv100, with limited documents per author, adding more exemplars may lead to diminishing returns. Overall, with 20 examples FT-E20, achieved the best performance overall with drop rates of 25.12\% for Blog50 and 10.14\% for IMDB62.

\noindent\textbf{LWF-based Exemplar Rehearsal. }Despite the FT-EK exemplar-based framework, the LWF-E10 and LWF-E2 models are standing in second place. Further testing with the LWF-E20 model, which uses twenty exemplars per author, showed strong results with drop rates of 26.38\% for Blog50 and 15.33\% for IMDB62, proofing the previous findings that increasing the number of exemplars can further enhance performance.

\noindent\textbf{Superiority of FT-EK. }\autoref{pd-analysis} illustrates the PD rates for the LWF-EK and FT-EK models across Blog50 and IMDb62 datasets using 6-session setups with K values of $2, 5, 10, \text{and } 20$. As depicted, both models exhibit lower accuracy drops when examples are used, indicating improved performance through the sessions. Notably, the FT-EK model, which outperforms due to its effectiveness in capturing new authors, shows superior results, especially evident in the CCAT50 and ArXiv100 datasets. This pattern is also supported by the results for CCAT50 shown in \autoref{conf-matrix}, with a low drop in the main diameter (high accuracy).

\section{Discussion}
\subsection{Closed-World AA. } The BERT model performance on case study datasets for closed-world AA highlights both strengths and limitations. As detailed in~\autoref{tab:performance_metrics}, the model excels with fewer authors and distinct writing styles, achieving 97.62\% accuracy on IMDB62. However, it shows a low performance with small number of documents per author, as shown by the 63.88\% accuracy on ArXiv100. The results on Blog50 and CCAT50 datasets ($\approx$80\% accuracy) indicate that overlapping styles pose challenges, and the expected difficulty with Blog1000 (45.56\% accuracy) further highlights limitations in handling larger datasets.  Overall, BERT is a powerful tool for AA, its performance is heavily influenced by the nature of the dataset. 

\begin{table}[t]
    \centering
    \caption{Closed-world AA Results on Test set using Accuracy Metric.}
    \begin{tabular}{|c|c|c|c|c|c|}
        \hline
        \textbf{Model} & \textbf{Blog50} & \textbf{IMDB62} & \textbf{CCAT50} & \textbf{Arxiv100} & \textbf{Blog1000} \\
        \hline
        BERT & 80.56& 97.62& 78.3 & 63.88& 45.56\\ \hline
    \end{tabular}
    \label{tab:performance_metrics}
\end{table}

\begin{table}[t]
    \centering
    \caption{Stylometric results using 6-session setups at CCAT50 dataset.}
    \begin{tabular}{|c|c|c|c|c|c|c|c|}
        \hline
        \textbf{Model} & s0 & s1 & s2 & s3 & s4 & s5 & PD $\downarrow$\\
        \hline
        FT+ &  87.6 &60.33 & 33.29 & 40.25 & 22.43 & 23.3& 64.3 \\ \hline
        Style FT+ & 85.2 & 55.67 & 35.14 & 35.0&25.44 & 27.20& 58 \\ \hline
        \hline
        FZ+   &    87.6 & 73& 62.57& 54.75& 48.67& 43.8& 43.8 \\ \hline
        Style FZ+ & 85.2& 71.17&61.14& 53.5 & 47.44 & 42.6& 42.6\\ \hline
        \hline
        FT\_E2     &   87.6 &    69.33& 61.29& 63.62& 55.33& 60.0& 27.6\\ \hline
        Style FT\_E2 & 85.2&72.83&62.71&61.88 & 53.11 & 62.4& \textbf{22.8}\\ \hline
        
    \end{tabular}
    \label{tab:style_results}
\end{table}

\subsection{Stylometric AA with CIL. }



However, the primary focus of this paper is to introduce CIL for AA. A comprehensive analysis of incorporating stylistic features specific to CIL AA is beyond the scope of this paper. Nonetheless, we present a preliminary exploration of such features within the CIL framework using the CCAT50 dataset. Similar to prior work~\cite{giglou-2021-profiling}, we combined the statistical feature representation of authors word usage probabilities with a BERT model as the backbone of the CIL approach. While the results reported in \autoref{tab:style_results} are not entirely consistent across all models, some interesting observations can be made. The \texttt{Style FT-E2} model achieved a PD rate of \textbf{22.8\%}, representing a 4.8\% improvement compared to the best FT-E2 model, although the initial performance of this model showed a slight drop compared to its counterparts. The \texttt{Style FZ+} model demonstrated a 1.2\% improvement in PD, while the \texttt{Style FT+} model exhibited a more significant improvement of 6.3\%. A possible explanation for the inconsistent performance of the style-incorporated models could be the fusion mechanism, which in our case is a simple concatenation of style and BERT representations. Moreover, the number of features extracted for style was limited to 100, which might not be sufficient, especially considering that the previous work involved only two groups of authors. We leave further exploration of such models for future work.

\section{FUTURE DIRECTIONS}
This section outlines potential areas for future research: 

\noindent\textbf{Source Code AA.} While we demonstrate CIL for AA using common datasets, it's crucial to recognize the broader scope of AA applications, including Source Code Attribution, which requires identifying the author of a code snippet. This area demands further exploration since the features extend beyond standard text. Notable datasets include those from the Google Code Jam (GCJ) competition and GitHub repositories, particularly for C and C++ code \cite{alsulami2017source, simko2018recognizing, abuhamad2018large}. These datasets emphasize the importance of incremental learning, as discussed in this paper. 

\noindent\textbf{Towards a fair comparison of evaluation metrics:}
As discussed in \ref{sec-eval}, commonly used metrics like average accuracy primarily focus on the overall accuracy achieved or lost at the end of incremental learning phases. However, these metrics may not be optimal as they fail to address the "stability-plasticity" dilemma, crucial for assessing the robustness and adaptability of approaches. Nevertheless, two approaches yielding nearly identical average scores can be misleading if one demonstrates high stability but lacks plasticity compared to another approach that balances both traits more effectively, making the latter more suitable. This phenomenon is particularly relevant in current CIL datasets where a significant portion of data is concentrated in the initial sessions. Approaches that merely memorize this initial data can achieve high average accuracy scores without demonstrating true adaptability over time. In our experiments with LWF and MAS, while their average performance and forgetting rates were similar, LWF exhibited a more balanced degree of stability-plasticity across datasets. 


\noindent\textbf{Moving toward FSCIL:} The optimal objective of incremental learning is to empower the learner to effectively manage the challenge posed by emerging authors with limited initial data, a common occurrence in practical applications. For instance, scholarly plagiarism detection systems must accommodate authors with sparse document counts initially, with the expectation that their document output will grow over time. Developing and refining an incremental learning framework tailored to handle these few-shot scenarios is crucial for advancing authorship attribution systems, thereby opening avenues for future research. A potential avenue involves enhancing models to better distinguish existing authors from each other and to streamline feature representations to accommodate new authors efficiently.

\section{Conclusions}
In this paper, we introduced CIL paradigm within the context of AA by providing a brief discussion of techniques commonly used in this domain and adapting some of them for several popular AA datasets. Consequently, the primary objective of this study is to reorient AA researchers' attention toward developing more effective AA systems that accommodate the continuous learning nature of AA. Finally, the results presented in this paper demonstrate the potential for adapting CIL in the field of AA. However, based on the performance of existing methods, there exists potential for further enhancement to achieve an effective model.

\bibliographystyle{IEEEtran}
\bibliography{ref}

\end{document}